\documentclass{elsart}
\usepackage{epsfig}

\renewcommand{\i}{i}
\newcommand{\re}{\mathop{\mathrm{Re}}}
\newcommand{\im}{\mathop{\mathrm{Im}}}
\newcommand{\tr}{\mathop{\mathrm{tr}}}

\renewcommand{\d}{d} 

\newcommand{\calD}{D}
\newcommand{\xit}{\xi_t}
\newcommand{\xis}{\xi^*}
\newcommand{\xist}{\xi_t^*}
\newcommand{\nut}{\nu_t}
\newcommand{\nus}{\nu^*}
\newcommand{\nust}{\nu_t^*}
\makeatletter
\renewcommand{\ps@copyright}{}
\makeatother

\begin{document}

\begin{frontmatter}



\title{Simulating spin-boson dynamics
with stochastic Liouville-von Neumann equations\protect\footnote{Dedicated
to the 60th birthday of Ulrich Weiss}}


\author{J\"urgen T. Stockburger}

\address{Institut f\"ur Theoretische Physik II,
Universit\"at Stuttgart
Pfaffenwaldring 57,
70550 Stuttgart, Germany}
\maketitle
\begin{abstract}
Based on recently derived exact stochastic Liouville-von Neumann
equations, several strategies for the efficient simulation of open
quantum systems are developed and tested on the spin-boson model.
The accuracy and efficiency of these simulations is verified for
several test cases including both coherent and incoherent dynamics,
involving timescales differing by several orders of magnitude. Using
simulations with a time-dependent field, the time evolution of
coherences in the reduced density matrix is investigated. Even in the
case of weak damping, pronounced preparation effects are found. These
indicate hidden coherence in the interacting system which can only be
indirectly observed in the basis of the reduced quantum dynamics.
\end{abstract}
\maketitle
\end{frontmatter}%
\section{Introduction}

The model of a two-level atom interacting with a quantum field or
many-particle reservoir has shown a remarkable persistence over decades
of progress in physics, appearing and reappearing in many guises in
different branches of condensed-matter physics after its `first life' in
quantum optics and magnetic resonance. It constitutes an idealized,
minimal model of quantum dynamics and thermodynamics of {\em open}\/
quantum systems. Through the explicit inclusion of a reservoir in the
model, it supports the discussion of an open quantum system without
resorting to speculative extensions of quantum mechanics.

The spin-boson problem \cite{chakr84,grabe84,legge87,weiss99}
is fairly well understood for most parameter
regimes relevant to solid-state physics, where the reservoir is
characterized by a smooth spectral density running up to a large
bandwidth cutoff, which enters the effective physics only as a
renormalization parameter. A rich body of theoretical work on the
spin-boson problem was developed in the context of macroscopic quantum
coherence \cite{legge86} and its realization in superconducting devices as
well as the physics of defect tunneling in solids \cite{esqui98}, to
name two early examples. These results are currently being reapplied and
extended in the discussion of performance limits of quantum computers,
since the two-level atom of the spin-boson problem can obviously be
identified with a qubit subject to a dephasing-inducing
environment \cite{makhl01}. Apart from this general consideration, the
spin-boson model also serves as a model for particular realizations,
e.g., the flux qubit \cite{chior03}, where physical parameters
can be given for the two-level system and its environment. Lastly, it
should be mentioned that an intricate formal link has recently been
pointed out \cite{fendl96,baur02} between the dynamics of the spin-boson
model and both dissipative transport of light particles\cite{schmi83} in
a periodic potential and transport of correlated electrons through a
barrier in a 1D conductor \cite{kane92}.

In the context of chemical physics, charge transfer and curve-crossing
problems form an important context for the spin-boson problem. Here the
understanding of the spin-boson dynamics seems less complete: The
bandwidth of the dissipative reservoir may be small, or its spectrum may
exhibit structured features relating to vibrational spectra. In addition
to the energy scales of the two-level system and a damping constant,
other parameters need to be considered. Moreover, the spin-boson
treatment of chemical phenomena is often considered only a first
approximation, since the linear-response assumption for the reservoir is
inherent in the model; a generalization to nonlinear, more complex
reservoirs is sometimes warranted. Since there is no mature theory of
non-linear reservoirs, the straightforward way to achieve such a
generalization is the formal inclusion of key degrees of freedom of the
reservoir, e.g., some prominent vibrational modes, in the open system.
But even the spin-boson problem itself is not fully understood absent
the scaling behaviour found in solid-state physics.

It is not to be expected that a thorough theoretical treatment of the
more complex models just indicated can be accomplished by analytic
methods alone; numerical simulations will become more important as
larger open systems are studied. Likewise, simulations are likely to
help in charting the remaining {\em terra incognita}\/ of the spin-boson
problem, e.g., the electron transfer problem in the inverted regime.
Future progress will place competing demands on numerical methods which
most established algorithms cannot meet at the same time: acceptable
scaling of computational complexity with {\em both} system size and time
interval simulated\cite{stock03}. Additionally, the method should take
into account the generally non-Markovian nature of quantum fluctuations.

The usual starting point for the theoretical description of an open
quantum system beyond perturbation theory, by now more or less
canonical, uses the path integral formalism, which allows the exact
treatment of quantum memory effects \cite{feynm63}. Since quantum
memory effects do not allow the path integral to be translated into a
simple differential equation of motion, this limits the choice of
theoretical and numerical approaches available. The direct evaluation
of path integrals using Monte Carlo sampling is a reliable method for
short to intermediate times \cite{egger94,leung95,muhlb03} but becomes
prohibitively expensive at long times due to the dynamical sign
problem. The recursive evaluation of a path integral using the
quasiadiabatic path-integral discretization (QUAPI) allows propagation
to arbitrarily long times, but needs computational resources which
grow very rapidly with increasing system size or memory time
\cite{makar94,sim96,thorw00}.

More recently, stochastic approaches have allowed a transition from the
path integral description to equations of motion
\cite{diosi98,stock98,strun99,stock99,stock01,stock02}. Similar to the
intuitive -- and computationally advantageous -- representation of
classical dissipative systems through the motion of a phase-space point
under the influence of thermal noise and friction, an open quantum
system can be described by the stochastic propagation of pure quantum
states \cite{diosi98,stock98,strun99,stock99,stock01,stock02,perci98,breue02}. Favourable scaling with
system size seems an {\em intrinsic} feature common to these approaches
-- the computational complexity of the open-system simulation as a function
of system size remains the same as that of simulating the same quantum
system without external interaction. The remainder of this article is
largely devoted to the question how the second objective, good scaling
for long-time dynamics, can be achieved in the presence of memory
effects. For tests of this performance aspect, the spin-boson system is
a prime candidate, since comparison results are readily available.

A summary of the formal description of dissipative quantum systems
through stochastic Liouville-von Neumann (SLN) equations given in Refs.
\cite{stock01,stock02} in Section \ref{sec:sln}. Section
\ref{sec:methods} discusses new material, giving details of numerical
approaches based on the SLN formalism. A series of numerical tests on
the free dynamics of the spin-boson model as well as simulation results
probing hidden coherence in the spin-boson model through pulsed external
fields are presented in Section \ref{sec:testres}, followed by a summary
of results and conclusions.  An appendix presents the derivation of a
mathematical result needed in Section \ref{sec:sln}.

\section{Stochastic Liouville-von Neumann equations for open quantum
  systems}
\label{sec:sln}

Although a connection between the influence functional formalism and
classical coloured noise in quantum dynamics was pointed out already in
the seminal work of Feynman and Vernon \cite{feynm63}, it has been put
to use in the theory and in simulations of open quantum systems
only recently \cite{diosi98,stock98,strun99,stock99,stock01,stock02}
 (see, however, \cite{kubo63,klein95}).
The key ideas of one such stochastic approach
to open quantum systems \cite{stock01,stock02} will be outlined in this
section using a more general model than the spin-boson model, commonly
known as the Caldeira-Legget model. The Caldeira-Legget model consists
of a one-dimensional potential model coupled to a thermal reservoir of
harmonic oscillators with a quasicontinuous distribution of frequencies,
\begin{equation}
  \label{eq:HCL}
  H = {p^2\over 2 m} + V(q) + \sum_{j}
  {m_j \omega_j^2\over 2}\left(x_j - {c_j\over m_j\omega_j^2} q\right)^2
  + {p_j^2\over 2 m_j}
\end{equation}

With the identification
\begin{equation}
  \label{qsigma}
  \hat{q} \to {q_0\over 2} \sigma_z ,
\end{equation}
where $q_0$ is a characteristic length scale, the spin-boson Hamiltonian
\begin{equation}
  \label{eq:SB}
  H_{\rm SB} = - {\hbar\Delta\over 2} \sigma_x
  + {\hbar\varepsilon\over 2} \sigma_z - {q_0\over 2} \sigma_z \sum_j c_j x_j
  + \sum_j {p_j^2\over 2 m_j}
  + {m_j \omega_j\over 2} x_j^2
\end{equation}
can be considered a truncation of the Caldeira-Legget model to
two localized states \cite{weiss87,legge87}.

The dynamics of the harmonic reservoir is usually of little interest;
moreover, due to the large size of the total system, only a reduced
description from which the oscillators have been eliminated is simple
enough to be practically treatable. Although Master equations have been
frequently been used with considerable success in weak-coupling
scenarios such as quantum optics \cite{walls94}, a formally exact
reduced formalism, usually necessary in a condensed-matter context, is
known only in path integral formalism. Feynman and Vernon \cite{feynm63}
have demonstrated that the path integral
\begin{eqnarray}
  \label{eq:CLPI}
  \rho(q_{\rm f},q'_{\rm f};t) &=&
  \int\limits \!\!\d q_{\rm i} \int\limits \!\!\d q'_{\rm i}
  \int\limits_{q_{\rm i}}^{q_{\rm f}} \!\!{ D}[q_1]
  \int\limits_{q'_{\rm i}}^{q'_{\rm f}} \!\!{ D}[q_2]\nonumber\\
  &&\times\exp\left({\i\over\hbar}( S_0[q_1] - S_0[q_2])\right)\nonumber\\
  &&\times F[q_1-q_2,(q_1+q_2)/2]\,
  \rho(q_{\rm i},q'_{\rm i};t_0)
\end{eqnarray}
is an exact expression for the time evaluation of the reduced density
matrix $\rho(q_{\rm f},q'_{\rm f};t)$ for an initial density matrix
which factorizes between $\rho(q_{\rm i},q'_{\rm i};t_0)$ and a thermal
density matrix of the harmonic reservoir. Here $S_0[q]$ denotes the
classical action functional associated with the one-dimensional
potential model. The influence functional
\begin{equation}
  \label{eq:IF1}
  F[x,r] = \exp\left(-{1\over\hbar} \Phi[x,r]\right)
\end{equation}
with
\begin{eqnarray}
  \label{eq:IF2}
  \Phi[x,r] &=& {1\over\hbar} \int\limits_{t_0}^t \!\!\d t' 
    \int\limits_{t_0}^{t'} \!\!\d t'' \,x(t') \nonumber \\
  && \times [\re L(t'-t'') x(t'')
  +2\i\im L(t'-t'') r(t'')] \nonumber \\
  &&+
  {\i\mu\over\hbar} \int\limits_{t_0}^t\!\!\d t'\, x(t') r(t') \;.
\end{eqnarray}
contains the full physics of the system-reservoir interaction and all
aspects of the reservoir dynamics which may be reflected in the system
dynamics. The kernel
\begin{equation}
  \label{eq:kernel}
  L(t) = {\hbar\over\pi} \int\limits_0^\infty \!\! \d\omega\,
  J(\omega)
  (\coth {\hbar\omega\beta\over2} \,\cos\omega t
  - \i\sin\omega t)
\end{equation}
describes free fluctuations of the reservoir's coupling coordinate,
themselves dependent on a spectral density $J(\omega)$ and inverse
thermal energy $\beta$. Typical spectral densities in the context of
solid-state physics are smooth in the relevant frequency regime, e.g.,
proportional to $\omega^3$ for bulk phonons or of the Ohmic form
$J(\omega) = \eta\omega$ for the low-energy excitations of a Fermi
liquid. In the case of the spin-boson model, a dimensionless constant
$\alpha = q_0^2 \eta/(2\pi\hbar)$ describes the strength of Ohmic
dissipation.

The last term in Eq. (\ref{eq:IF2}) has the form of a potential
modification, which reflects the fact that the system-reservoir
interaction given in Eq.(\ref{eq:HCL}) eliminates any quasistatic
response of the reservoir; it relates to dynamic reservoir's dynamic
response function $\chi_R(t-t') = -2\Theta(t-t') \im L(t-t')/\hbar$
through $\mu = \int_0^\infty\d\tau \chi_R(\tau)$. This potential term
has no effect for the spin-boson model, from $q^2\equiv q_0^2/4$ one
finds $\mu xr = (\mu/2) (q_1^2 - q_2^2) = 0$.

The path integral expression (\ref{eq:CLPI}) has the distinct advantage
of providing an exact description of the system-plus-reservoir dynamics
which is reduced to only the system degree of freedom. However, the
propagator for the reduced density matrix described by Eq.
(\ref{eq:CLPI}) is {\em not}\/ associative due to the fact that the
exponent of $F[x,r]$ contains a {\em double}\/ time integral, i.e.,
there are memory effects which make quantum amplitudes a non-local
functional on the path space.

At the price of introducing a further functional integral over
auxiliary function spaces of complex functions $\xi(t)$ and $\nu(t)$, this
non-locality can be lifted, representing the influence functional as a
weighted average of quantum amplitudes containing a time-local action
functional
\begin{eqnarray}
  \label{eq:Hubbard}
  F[x,r] &=& \int\calD^2[\xi] \int\calD^2[\nu] W[\xi,\xi^*,\nu,\nu^*]\\
  &&\times\exp\left({\i\over\hbar} \int\limits_{t_0}^t \d t'\xi(t') x(t')
    + \i \nu(t')r(t')\right)
  \nonumber\\
  &&\times\exp\left( -{\i\mu\over\hbar}\int\limits_{t_0}^t \d t'
    x(t')r(t') \right)
  \nonumber
\end{eqnarray}
with a suitable Gaussian functional $W[\xi,\xi^*,\nu,\nu^*]$. Because
$W$ is normalized, it may be interpreted as a probability density of
real noise fluctuations $(\xi+\xi^*)/2$, $(\xi-\xi^*)/(2\i)$,
$(\nu+\nu^*)/2$ and $(\nu-\nu^*)/(2\i)$. It is partly characterized by
the noise correlation functions
\begin{eqnarray}
  \langle \xi(t) \xi(t') \rangle_W &=& {\rm Re} L(t-t')
  \label{eq:xiauto}\\
  \langle\xi(t)\nu(t')\rangle_W &=& (2\i /\hbar) \Theta(t-t') \im
    L(t-t')\nonumber \\
  &&= -\i\chi_R(t-t') \label{eq:xinu}\\
  \langle\nu(t)\nu(t')\rangle_W  &=& 0  \label{eq:nuauto}
\end{eqnarray}
required for the identification of Eqs. (\ref{eq:Hubbard}) and
(\ref{eq:IF1}). $W$ is also characterized by the correlations
$\langle\xi(t)\xi^*(t')\rangle$, $\langle \xi(t)\nu^*(t')\rangle$ and
$\langle\nu(t)\nu^*(t')\rangle$, which do not enter the physical result
obtained upon stochastic averaging.

The immediate benefit of the stochastic construction (\ref{eq:Hubbard})
becomes clear if the order of the integrations over $(q,q')$ and
$(\xi,\nu)$ is interchanged: for any specific functions $\xi(t)$ and
$\nu(t)$ the path-integral dynamics can be translated into the
Schr\"odinger picture in the usual way. Observing that the exponents in
Eq. (\ref{eq:Hubbard}) can be identified as action functionals
associated with a time-dependent potential, we immediately recover
the equation of motion
\begin{equation}
  \label{eq:slnlin}
  \i\hbar\dot\rho = [ H_0,\rho ]_- - \xi(t) [q,\rho]_- 
  + {\mu\over 2} [q^2,\rho]_-
  - {\hbar\over 2}\nu(t) [q,\rho]_+ ,
\end{equation}
a stochastic Liouville-von Neumann (SLN) equation. The presence of both
complex parameters and an anticommutator makes this equation describe a
non-unitary propagation of individual samples, allowing $\rho$ to
acquire a non-hermitean component. After stochastic averaging, however,
any non-hermitean parts of $\rho$ vanish. Similarly, the trace of $\rho$
varies between individual samples, but remains equal to unity on average.

This diffusive spread of the sample trace can be reduced or eliminated
by introducing a `guide trajectory' (to be specified later) in the
anticommutator term,
\begin{equation}
  \label{eq:slnguide}
  \i\hbar\dot{\hat\rho} = [ H_0,\hat\rho ]_- - \xi(t) [q,\hat\rho]_-
  + {\mu\over 2} [q^2,\hat\rho]_-
  - {\hbar\over 2}\nu(t) [q-\bar r_{t},\hat\rho]_+,
\end{equation}
where
\begin{equation}
  \label{eq:guidefctr}
  \rho(t) = \exp\left(i\int\limits_0^t \d t' \nu(t') \bar r_{t'}\right)
  \hat\rho(t)
\end{equation}
relates $\hat\rho$ to the solution of the original equation
(\ref{eq:slnlin}) for arbitrary $\bar r_t$. It is to be noted that
Eqs. (\ref{eq:slnguide}) and (\ref{eq:guidefctr}) do not set the stage
for an approximate expansion; together they form an exact identity.
The guide trajectory may depend on the noise forces in the form of a
functional $\bar{r}_t[\xi(t'),\nu(t')]$, where the following properties
will be assumed:
\begin{eqnarray}
\label{eq:causxi}
&&  {\delta \bar{r}_t \over \delta \xi(t')} = 0\,; \; t'\geq t \\
\label{eq:causnu}
&&  {\delta \bar{r}_t \over \delta \nu(t')} = 0\,; \; t'\geq t \\
\label{eq:anlyxic}
&&  {\delta \bar{r}_t \over \delta \xi^*(t')} = 0\,; \; \mbox{any } t' \\
\label{eq:anlynuc}
&&  {\delta \bar{r}_t \over \delta \nu^*(t')} = 0\,; \; \mbox{any } t' .
\end{eqnarray}
In words, the functional $\bar r_t[\xi(t'),\nu(t')]$ is required to be
{\em causal}\/ and {\em analytic}.

Under these conditions, the exponential factor in Eq.
(\ref{eq:guidefctr}) can be absorbed into the probability measure by
re-defining the centre of the Gaussian functional
$W[\xi,\xi^*,\nu,\nu^*]$ (see Appendix). This leads to new noise forces
with same variance but dynamically shifted mean values,
\begin{eqnarray}
\xi_t(t') &=& \xi(t') - \i\int\limits_0^t \!\!\d s\,
 \langle\xi(t')\nu(s)\rangle \bar{r}(s) \nonumber\\
&& = \xi(t') + \int\limits_0^t \!\!\d s\, \chi_R(t'-s) \bar{r}(s)
\label{eq:xishift} \\
\xi^*_t(t') &=& \xi^*(t') - \i\int\limits_0^t \!\!\d s\,
 \langle\xi^*(t')\nu(s)\rangle \bar{r}(s)
\label{eq:xistarshift} \\
\nu_t(t') &=& \nu(t')
\label{eq:nushift} \\
\nu^*_t(t') &=& \nu^*(t') - \i \int\limits_0^t \!\!\d s\,
\langle\nu^*(t')\nu(s)\rangle \bar{r}(s)
\label{eq:nustarshift}
\end{eqnarray}
Dropping the subscript $t$ after the transformation has been performed,
we obtain the SLN equation
\begin{eqnarray}
  \label{eq:slnnorm}
  \i\hbar\dot{\hat\rho} &=& [ H_0,\hat\rho ]_-
  - \left(\xi(t) - \int_{t_0}^t \d
  t' \chi_R(t-t') \bar{r}_{t'} \right)[q,\hat\rho]_- \\
  &&
  + {\mu\over 2} [q^2,\hat\rho]_-
  - {\hbar\over 2}\nu(t) [q-\bar r_{t},\hat\rho]_+
  \nonumber
\end{eqnarray}
to be averaged with the {\em original} probability density
$W[\xi,\xis,\nu,\nus]$ without the exponential factor of
Eq. (\ref{eq:guidefctr}).

A natural choice for $\bar{r}_t$ relates it to the dynamics on the time
interval $[0,t]$,
\begin{equation}
  \label{eq:rbardyn}
  \bar{r}_t = {\tr q\rho(t)\over \tr\rho(t)} = \tr q\hat\rho(t) .
\end{equation}
Since this makes Eq. (\ref{eq:slnnorm}) conserve the trace of $\hat\rho$
for each sample, we call Eq. (\ref{eq:slnnorm}) a {\em normalized}\/ SLN
equation for a guide trajectory defined by Eq. (\ref{eq:rbardyn}).
Compared to the linear SLN equation (\ref{eq:slnlin}) the normalized
version allows a more efficient stochastic averaging as well as a clear
interpretation of the classical limit of Eq. (\ref{eq:slnnorm})
\cite{stock01,stock02}.

The implementation of a numerical simulation algorithm based on the
normalized SLN equation is now relatively straightforward. For a
numerical construction of the noise fluctuations, $\xi$ is decomposed
into a purely real $\xi^{(l)}$ with a long autocorrelation timescale,
which is uncorrelated to all other noise forces, and a short-range
complex part $\xi^{(s)}$, with correlations
\begin{eqnarray}
  \label{eq:xilxis}
  \langle\xi^{(l)}(t)\xi^{(l)}(t')\rangle &=&  \re L(t-t') \\
  \langle\xi^{(s)}(t)\xi^{(s)}(t')\rangle &=&  0 \\
  \langle\xi^{(s)}(t)\nu(t')\rangle &=& - \i  \chi_R(t-t') \\
  \langle\xi^{(s)}(t)\nu^*(t')\rangle &=& 0 \\
  \langle\nu(t)\nu(t')\rangle &=& 0 .
\end{eqnarray}
Here the correlations of $\xi^{(s)}$ and $\nu$ decay rapidly for
$\omega_c |t-t'| \gg 1$, while correlations of $\xi^{(l)}(t)$ persist
up to the thermal timescale $\hbar\beta$.

These noise fluctuations are generated numerically with great efficiency
by filtering white noise with suitable integral operators, whose kernels
have spectra which multiply to yield the spectra of $L(t-t')$ and
$\chi_R(t-t')$. The necessary convolution operations can be performed
with ease using a fast Fourier transform algorithm.  This construction
also determines the non-physical correlations
$\langle\xi^{(s)}(t)\xi^{(s)*}(t')\rangle$ and
$\langle\nu(t)\nu^*(t')\rangle$.

The nonlinearity of the numerically simulated dynamics can be reduced by
observing that the solution of the {\em simplified} SLN equation
\begin{eqnarray}
  \label{eq:slnnum}
  \i\hbar\dot{\rho} &=& [ H_0,\rho ]_- - \left(\xi(t) - \int_{t_0}^t \d
    t' \chi_R(t-t') \bar{r}_{t'} \right)[q,\rho]_- \\
  &&
  + {\mu\over 2} [q^2,\rho]_-
  - {\hbar\over 2}\nu(t) [q,\rho]_+
  \nonumber
\end{eqnarray}
immediately yields a solution of the nonlinear SLN equation
(\ref{eq:slnnorm}) through the relation
\begin{equation}
  \label{eq:ansatz}
  \hat\rho(t) = {1\over \tr\rho(t)} \rho(t) .
\end{equation}

A further simplification of Eq. (\ref{eq:slnnum}) lies in the
factorizing ansatz $\rho = |\psi_1\rangle\langle \psi_2|$, which reduces
Eq. (\ref{eq:slnnum}) to the two Schr\"odinger equations

\begin{eqnarray}
  \i\hbar|\dot\psi_1\rangle
  &=& H_0|\psi_1\rangle - \left(\xi(t) - \int_{t_0}^t \d
    t' \chi_R(t-t') \bar{r}_{t'} \right) q |\psi_1\rangle \nonumber\\
  \label{eq:schroe1}
  && + {\mu\over 2} q^2 |\psi_1\rangle
  -
  {\hbar\over 2}\nu(t) q |\psi_1\rangle \\
  \i\hbar|\dot\psi_2\rangle
  &=& H_0|\psi_2\rangle - \left(\xis(t) - \int_{t_0}^t \d
    t' \chi_R(t-t') \bar{r}_{t'}^* \right) q |\psi_2\rangle \nonumber \\
  \label{eq:schroe2}
  && + {\mu\over 2} q^2 |\psi_2\rangle
  +
  {\hbar\over 2}\nus(t) q |\psi_2\rangle
\end{eqnarray}

Apart from the satisfaction of arriving at a conceptually simple result
from an involved mathematical transformation, this reflects very
positively on the scaling of stochastic simulations: For a single noise
sample, the numerical effort of propagating the system scales with
system size exactly the same way as for a conservative system. Moreover,
it is interesting to note that Eqs.  (\ref{eq:schroe1}) and
(\ref{eq:schroe2}) capture all effects of the system-reservoir
interaction, including quantum correlations and entanglement between
system and reservoir, through the addition of {\em single-particle}
operators to the Hamiltonian $H_0$. The description of quantum
dissipation through single-particle operators strongly suggests that a
wide range of approximations to conservative quantum systems (e.g.,
formalisms suitable for large atoms or small molecules) can be combined
with this stochastic approach. Larger and more complex quantum systems,
which can currently be simulated only under the assumption of energy
conservation, are likely to become accessible to an open-system approach
using SLN equations. There is an attractive feature in in the SLN
approach which may recommend it for complex systems even in the case of
very weak damping, where perturbative methods are valid: Whereas
Born-Markov perturbative approaches such as Redfield equations define
damping in terms of system {\em and}\/ reservoir characteristics, often
making reference to an exact diagonalization of the system, the SLN
noise terms are constructed from the free {\em reservoir} dynamics
alone; no characteristics of the free system dynamics enter the formal
description of dissipation.

\section{Optimized stochastic simulation methods}
\label{sec:methods}

For the transformed noise forces given in Eqs.(\ref{eq:causxi}) --
(\ref{eq:anlynuc}) the linear combinations $\xi_t\pm\xi^*_t$ and
$\nu_t\pm\nu^*_t$ are no longer purely real or purely imaginary. The
transformed integration measure is formally Gaussian, but the
integration, even when written over the real components of $\xi$ and
$\nu$, is shifted in the complex domain. However, the stochastic
simulation of Eq.  (\ref{eq:slnnum}) must use real-valued components for
these shifted quantities, i.e., the numerical simulation of Eq.
(\ref{eq:slnnum}) operates on an analytic continuation of the exact
result.  Although this continuation is known to be free of singularities
in important cases such as the classical limit, the weak-coupling limit
and any harmonic system, the spin-boson model (at low temperature)
appears to be a counterexample where the simulation becomes
unstable after returning excellent results up to a well-defined time
threshold $t_c$, which is related to the growth of the non-hermitean
part of $\hat\rho$.  Beyond the threshold a sudden onset of peak-like
artefacts and other systematic errors is observed, in close similarity
to stability problems found in the Master equations for the positive $P$
function, a quantum optical phase space function \cite{schac91,gardi00}.
As in the latter example, SLN simulations using Eq.  (\ref{eq:slnnum})
are highly accurate for times below the threshold time. In the light of
these facts two natural ways to extend the power of the SLN methodology
are (a) increasing the timescale $t_c$ and (b) finding consistent
methods to `reshuffle' the ensemble to a healthy state before artefacts
taint the data.

\subsection{Alternative noise spectra}

In the scaling limit any relevant frequency $\Omega$ of the dissipative
dynamics lies within the band of reservoir frequencies,
$\Omega\ll\omega_c$, i.e., the complex noise forces, which drive the
system away from normalized hermitean states, act on the system at
resonance. This can be changed if the potential term in the influence
functional is formally included in the stochastic construction, which
leads to modified cross-correlations for the noise fluctuations
$\xi^{(s)}$ and $\nu$,
\begin{equation}
  \label{eq:ndnoise}
  \langle\xi^{(s)}(t)\nu(t')\rangle = -\i\chi_{\rm R}(t-t') +\i\mu
  \delta(t-t') .
\end{equation}
Instead of a noise spectrum extending roughly over the interval
$[-\omega_c,\omega_c]$, we are now dealing with a white noise spectrum
that has a {\em gap} in the same interval. The resonant driving of the
system by complex noise is thus strongly reduced,  typically increasing
the critical time $t_c$ by one or more orders of magnitude. With the
proviso that increments of the white noise parts of $\xi$ and $\nu$ be
interpreted in the sense of a Stratonovich stochastic integral, the
steps taken in the derivation of Eqs. (\ref{eq:slnnorm}) and
(\ref{eq:slnnum}) can be reiterated unchanged. This yields
\begin{eqnarray}
  \i\hbar\dot{\rho} &=& [ H_0,\rho ]_- - \left(\xi(t) - \int_{t_0}^t \d
    t' \chi_R(t-t') \bar{r}_{t'} + \mu \bar{r}_t \right) [q,\rho]_-
  \nonumber \\
  && - {\hbar\over 2}\nu(t) [q,\rho]_+ .
  \label{eq:slnstratonovic}
\end{eqnarray}
as a modified version of the simplified SLN equation (\ref{eq:slnnum}),
which has the additional advantage of having only translationally
invariant dissipation terms.

\subsection{Collective normalization of subensembles}

Next we consider the case of a subensemble of $N$ samples $\rho^{(j)}(t)$,
each governed by an independent set of noise forces $\xi^{(j)}(t)$,
$\nu^{(j)}(t)$. In this case, a set of guide trajectories
$\bar{r}_t^{(j)}$ can be given which does not conserve the trace of each
sample, but only their sum. Again, guide trajectories enter an
exponential factor distinguishing normalized and unnormalized states,
but here a single factor is used for different noise realizations,

\begin{equation}
  \label{eq:collnorm}
  \rho^{(j)}(t) = \prod_{k=1}^N \exp\left(\i \int_0^t \d t'
  \nu^{(k)}(t') \bar{r}_t^{(k)}\right) \hat\rho^{(j)}(t')
\end{equation}
with
\begin{equation}
  \label{eq:rbarj}
  \bar{r}_t^{(k)} = {\tr q \rho^{(k)}(t)
    \over \sum_{l=1}^N \tr \rho^{(l)}(t)}
   = {1\over N} \tr q \hat\rho^{(k)}(t)
\end{equation}
For initially normalized states, this keeps the sum over traces
identical to $N$.

Again, the exponential factors in Eq. (\ref{eq:collnorm}) must be
absorbed into the probability measure through suitable shifts of the
noise fluctuations. Because the (unshifted) noises acting on different
members of the subensemble are uncorrelated, and because terms with
different $\nu^{(k)}$ in Eq. (\ref{eq:collnorm}) factorize, the
shifts of noise fluctuations $(\xi^{(j)}, \xi^{(j)*}, \nu^{(j)},
\nu^{(j)*})$ and $(\xi^{(k)}, \xi^{(k)*}, \nu^{(k)}, \nu^{(k)*})$ with
$j\neq k$ are {\em independent} of each other. Thus the derivation
presented in the Appendix applies also in this case. Eq.
(\ref{eq:xishift}) is only trivially modified for collectively
normalized samples,
\begin{equation}
  \label{eq:collxishift}
  \xi_t^{(j)}(t') =\xi^{(j)}(t')
  + \int\limits_0^t \!\!\d s\, \chi_R(t'-s) \bar{r}_t^{(j)}(s) .
\end{equation}
The corresponding normalized SLN equation reads
\begin{eqnarray}
  \nonumber
  \i\hbar\dot{\hat\rho}^{(j)} &=& [ H_0,\hat\rho^{(j)} ]_-
  - \left(\xi^{(j)}(t)
    - \int_{t_0}^t \d t' \chi_R(t-t') \bar{r}^{(j)}_{t'} \right)
  [q,\hat\rho^{(j)}]_- \\
  \label{eq:slncolnorm}
  &&
  + {\mu\over 2} [q^2,\hat\rho]_-
  - {\hbar\over 2}\nu^{(j)}(t) [q,\hat\rho^{(j)}]_+
  + \sum_{k=1}^N \hbar\nu^{(k)}(t) \bar{r}^{(k)}_t \hat\rho^{(j)} .
\end{eqnarray}
Again, there is a simplified SLN equation suitable for numerical
simulation,
\begin{eqnarray}
  \label{eq:slncolnum}
  \i\hbar\dot\rho^{(j)} &=& [ H_0,\rho^{(j)} ]_-
  - \left(\xi^{(j)}(t)
    - \int_{t_0}^t \d t' \chi_R(t-t') \bar{r}^{(j)}_{t'} \right)
  [q,\rho^{(j)}]_- \nonumber \\
  &&
  + {\mu\over 2} [q^2,\rho]_-
  - {\hbar\over 2}\nu^{(j)}(t) [q,\rho^{(j)}]_+ ,
\end{eqnarray}
related to the normalized form (\ref{eq:slncolnorm}) by
\begin{equation}
  \label{eq:ansatzcol}
  \hat\rho^{(j)}(t) = {N\over \sum_{k=1}^N \tr\rho^{(k)}(t)}
  \rho^{(j)}(t) .
\end{equation}
It is evident from Eq.  (\ref{eq:rbarj}) that $\bar{r}_t^{(j)}$ is of
order $1/N$, i.e., the shift (and the corresponding nonlinearity of the
equation of motion) vanishes for large $N$; the threshold time $t_c$ is
thus further extended. The linear SLN equation (\ref{eq:slnlin}) is
recovered from (\ref{eq:slncolnum}) in the limit $N\to \infty$.

\subsection{Separation of friction and thermal time scales}
\label{sec:blocks}

\begin{figure}[tb]
  \begin{center}
    \epsfxsize=0.7\columnwidth
    \epsffile{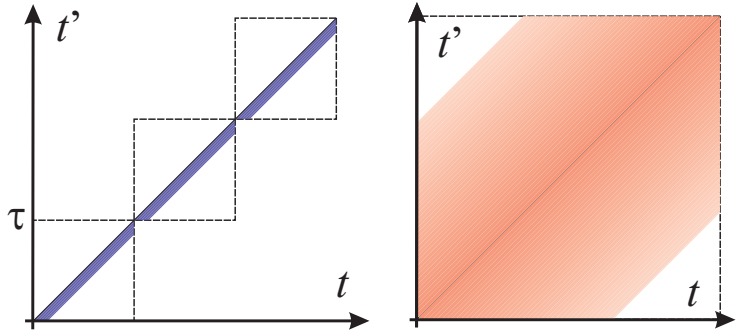}
    \caption[]{Schematic representation of block factorization of
      short-time correlations (left) vs. global reach of long-time
      correlations (right) (see subsection \ref{sec:blocks}).}
    \label{fig:blocks}
  \end{center}
\end{figure}

The split of $\xi$ into the long- and short-range components $\xi^{(l)}$
and $\xi^{(s)}$ allows the numerical solution of SLN equations using a
two-stage averaging procedure, which separates the difficulty of dealing
with long-range correlations from the problems incurred through complex
driving forces. In a primary averaging stage (inner sampling loop),
samples are drawn for $\xi^{(s)}$ and $\nu$, while $\xi^{(l)}$ will be
changed only in the secondary averaging stage (outer sampling loop). For
simulation parameters yielding a sufficiently large threshold time,
$\omega_c t_c \gg 1$, the short-time noise correlation
$\langle\xi(t)\nu(t')\rangle$ can be block factorized on intervals of
width $\tau < t_c$, as indicated in Fig. \ref{fig:blocks} (left). Under
the condition $\omega_c \tau \gg 1$ this leaves the effective support of
the correlator $\langle\xi(t)\nu(t')\rangle$ virtually untouched.

In the primary averaging, the dynamical simulation can thus be
partitioned over finite time intervals, accumulating and then drawing
new Hermitean samples at the end of each interval. This procedure does
not disturb the long-time correlations since all samples drawn for the
primary averaging evolve subject to the same realization of $\xi^{(l)}$.
In the secondary averaging, the entire procedure is repeated for a
sufficient number of long-range fluctuations $\xi^{(l)}$, which remain
unconstrained in the $t$,$t'$ plane. Their effective support forms a
diagonal band of width $\propto \hbar \beta$ (Fig.  \ref{fig:blocks},
right).

\section{Tests and results for the spin-boson model}
\label{sec:testres}

Numerical tests and applications of the simulation strategies described
in Section \ref{sec:methods} are discussed in the following; a
combination of all three strategies is used unless indicated otherwise.
The underlying SLN equations of motion for the spin-boson model are as
follows: The linear SLN equation reads
\begin{equation}
  \label{eq:SB:SLN:lin}
  \i\hbar\dot\rho = - {\hbar\Delta\over 2} [\sigma_x,\rho]_-
  + {\hbar\varepsilon\over 2} [\sigma_z,\rho]_-
  - \xi(t) [\sigma_z,\rho]_- - {\hbar\nu\over 2} [\sigma_z,\rho]_+
  .
\end{equation}
Absent any physical context defining a characteristic length $q_0$,
$\sigma_z$ is identified with the position operator $q$.  The normalized
and simplified SLN equations are
\begin{eqnarray}
  \i\hbar\dot{\hat\rho} &=& - {\hbar\Delta\over 2} [\sigma_x,\hat\rho]_-
  + {\hbar\varepsilon\over 2} [\sigma_z,\hat\rho]_-
  \nonumber \\
  &&- \left(\xi(t) - \int_{t_0}^t \d t' \chi_R(t-t') \bar{\sigma}_{t'} \right)
  [\sigma_z,\hat\rho]_-
  \nonumber \\
  && - {\hbar\nu\over 2} [\sigma_z-\bar\sigma_t,\hat\rho]_+
  \label{eq:SB:SLN:norm}
\end{eqnarray}
(with $\bar\sigma_t = \tr\sigma_z\hat\rho$) and
\begin{eqnarray}
  \i\hbar\dot{\rho} &=& - {\hbar\Delta\over 2} [\sigma_x,\rho]_-
  + {\hbar\varepsilon\over 2} [\sigma_z,\rho]_-
  \nonumber \\
  && - \left(\xi(t) - \int_{t_0}^t \d t' \chi_R(t-t') \bar{\sigma}_{t'} \right)
  [\sigma_z,\rho]_-
  \nonumber \\
  && - {\hbar\nu\over 2} [\sigma_z,\rho]_+
  \label{eq:SB:SLN:simp}
\end{eqnarray}
(with $\hat\rho = \rho/\tr\rho$), respectively. Throughout the
following, Ohmic dissipation with an algebraic cutoff
\begin{equation}
  \label{eq:ohmalg}
  J(\omega) = {\eta \omega \over \left(1+{\omega^2\over\omega_c^2}\right)^2}
\end{equation}
corresponding to an exponentially decaying response function
\begin{equation}
  \label{eq:chiexp}
  \chi_{\rm R}(t) = \mu\omega_c^2 t \exp(-\omega_c t)
\end{equation}
is used.

\subsection{Coherent dynamics: free oscillations}

\begin{figure}[tb]
  \begin{center}
    \epsfxsize=0.6\columnwidth
    \epsffile{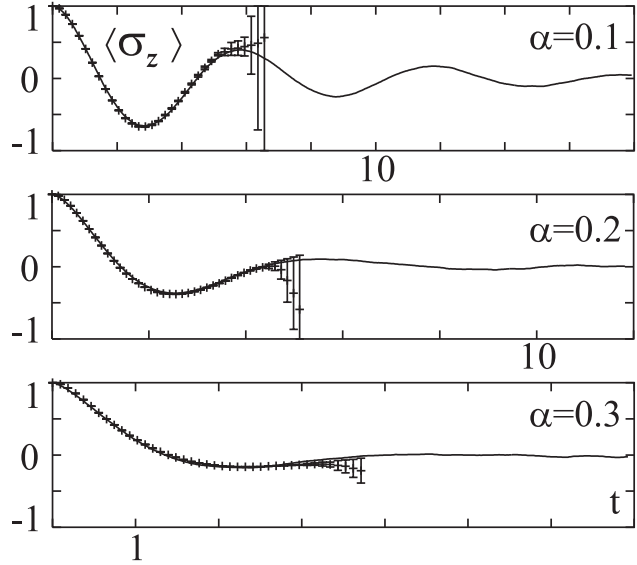}
    \caption[]{Comparison of SLN simulation data (lines) and
      path-integral Monte Carlo data (symbols) for coherent spin-boson
      dynamics at zero temperature. The time $t$ is scaled by the
      renormalized matrix element $\Delta_{\rm r}$ (see text).}
    \label{fig:coherent}
  \end{center}
\end{figure}

The coherent dynamics of the spin-boson model at moderate damping,
$\alpha < 1/2$, and low temperature are theoretically well understood
and numerically accessible using other methods, thus providing a
suitable test case to verify the accuracy and stability of the SLN
approach. Using data obtained through an established path-integral Monte
Carlo (PIMC) method for comparison \cite{egger94}, the power and
efficiency of the SLN approach in the region of coherent dynamics at
zero temperature is demonstrated. Fig. \ref{fig:coherent} shows the
time evolution of the expectation value $\langle \sigma_z\rangle$ of a
symmetric system from an initially factorized state.

In the SLN approach the dynamics can be accurately simulated over many
oscillation periods; empirical data suggest that the growth of
statistical errors saturates to a constant value for increasing $t$,
allowing the simulation to be continued to arbitrarily long times.
This overcomes an inherent restriction of the PIMC approach, which
suffers from the so-called dynamical sign problem: The PIMC approach
mainly employs the superposition principle to obtain quantum amplitudes.
The resulting averaging over highly oscillatory functions results in a
signal-to-noise ration of the MC simulation which degrades exponentially
with growing $t$, leading to near-divergent error estimates after about
one oscillation period. For very strong system-reservoir coupling,
however, quantum phase factors are suppressed, and the PIMC approach
gains in efficiency.

The computer times of the SLN and PIMC simulations shown in Fig.
\ref{fig:coherent} are comparable; they vary between 40 h and 210 h on a
commodity CPU (Athlon, 1200 MHz). The SLN simulations, whose statistical
errors are comparable to the linewidth, use between 10 and 20
factorization blocks. 
The time axis is scaled by the renormalized matrix element $\Delta_{\rm
  r} = \Delta (\Delta/\omega_c)^{\alpha/(1-\alpha)}$. In the SLN
simulations a high cutoff $\omega_c = 20 \Delta$ was used while a lower
cutoff $\omega_c = 6 \Delta$ was used in the PIMC simulation to avoid
ergodicity problems.

\subsection{Incoherent relaxation: long-time dynamics}

\begin{figure}[tb]
  \begin{center}
    \epsfxsize=0.6\columnwidth
    \epsffile{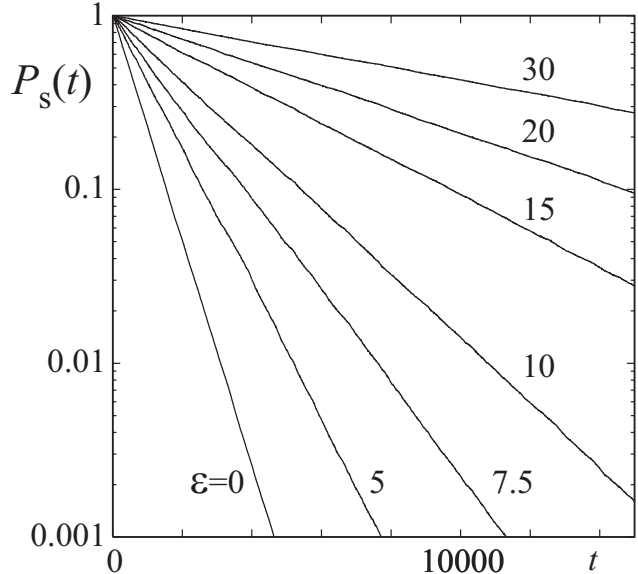}
    \caption[]{Exponential relaxation of the asymmetric spin-boson
    system. All curves agree with analytic nonadiabatic rates within a
    few percent. The time $t$ is given in units of $\Delta$.}
    \label{fig:rate}
  \end{center}
\end{figure}

Simulating the slow exponential decay of localized populations in a
biased spin-boson model, as it occurs, e.g., in electron transfer
reactions, is chosen as another test case.  For parameter regions with
non-trivial transient behaviour, an explicit simulation of
time-dependent population numbers can offer advantages over simulations
linking rates to thermal flux correlation functions \cite{muhlb03}. The
population difference $P(t)$ is chosen as a dynamical observable; $P(t)$
is formally defined as the time-dependent expectation value
$\sigma_z(t)$ for a two-state system starting from the factorized
initial state with
\begin{equation}
  \label{eq:rho0}
  \rho(t=0) = \left(\begin{array}{cc}
      1&0\\0&0
    \end{array}\right) 
\end{equation}
and the reservoir in thermal equilibrium.
For rate calculations, it is essential that $P(t)$ can be simulated over
a long enough time interval to be directly fitted to an exponential
function. Since $P(t)$ decays towards a nonzero value in a biased
system, it is advantageous to symmetrize it under a reversal of the bias
sign, $\varepsilon \to -\varepsilon$, i.e., study
\begin{equation}
  \label{eq:ps}
  P_s(t) = {1\over 2} (P_{\varepsilon}(t) + P_{-\varepsilon}(t)) .
\end{equation}
Because changing the sign of $\varepsilon$ is the same as
interchanging the eigenstates of $\sigma_z$, this can be expressed as a
single expectation value with an {\em antisymmetric}\/ initial state
\begin{equation}
  \label{eq:ps1}
  P_s(t) = \langle\sigma_z(t)\rangle\;, \quad
  \rho(t=0) = {1\over 2} \left(\begin{array}{lr}
      1&0\\0&-1
    \end{array}\right) .
\end{equation}
Using this equivalent definition of the symmetrized expectation value
$P_s(t)$ as the basis of numerical simulations,the data presented in
Fig. \ref{fig:rate} is obtained. Parameters are $\alpha = 0.1$,
$\omega_c/\Delta = 100$, $k_{\rm B}T/\hbar\Delta = 10$ and $\varepsilon$
as indicated in the figure. Since the simulation time is many
orders of magnitude longer than the inverse of $\omega_c$, the
relatively large number of 300 factorization blocks may be used. In
spite of the large number of time steps, the computation of a single
curve in Fig. \ref{fig:rate} took only about 15 min.

For rates varying over more than an order of magnitude, a remarkably
stable simulation result for exponential decay is found, extending over
many time constants. Given the fact that these curves result from the
averaging of {\em coherently} propagated samples, this is truly
remarkable. In the light of this successful example of a highly
efficient simulation, it is to be expected that more elaborate
simulations using this technique will be powerful enough to generate
benchmark data for electron transfer rates which can be used to check
theoretical approximations.

\subsection{Rapid dephasing: short-time dynamics}
\begin{figure}[tb]
  \begin{center}
    \epsfxsize=0.6\columnwidth
    \epsffile{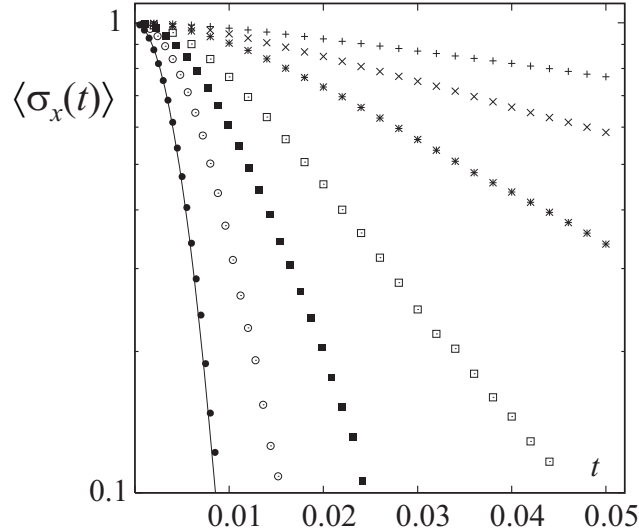}
    \caption[]{Rapid decoherence of a symmetric superposition of localized
      states with a factorized initial density matrix and
      $\omega_c/\Delta = 100$, $k_{\rm B}T/\hbar\Delta = 20$ and with
      $\alpha$ varying from $5$ (symbol `$\bullet$', curve at left) to
      $0.05$ (symbol `$+$', at top). The time $t$ is given in units of
      $\Delta$.}
    \label{fig:decoh}
  \end{center}
\end{figure}

The dephasing of a coherent superposition of quantum states is one of
the fastest processes in spin-boson dynamics. It is mainly governed by
the {\em interaction} part of the Hamiltonian (\ref{eq:SB})
\cite{braun01}. After preparing the system in an eigenstate of
$\sigma_x$, again with a factorizing initial condition, dephasing
expresses itself through the decay of the off-diagonal elements of the
density matrix, here probed by measuring $\langle \sigma_x(t)\rangle$.
The results given in Fig.  \ref{fig:decoh} are for different values
$\alpha = 5$, $2$, $1$, $0.5$, $0.2$, $0.1$ and $0.05$ of the
dissipation constant (different symbols, from bottom left to top right).

As outlined in Ref. \cite{braun01}, a transition from a rapid Gaussian
decay towards a slower, more complicated decay is observed when varying the
damping constant from strong to weak damping. Excellent agreement
between the analytic strong-coupling result (solid line) and simulation
(filled dots) is observed. The statistical errors are less than $\pm
0.02$.

\subsection{Preparation effects: probing hidden coherence}

\begin{figure}[tb]
  \begin{center}
    \epsfxsize=0.6\columnwidth
    \epsffile{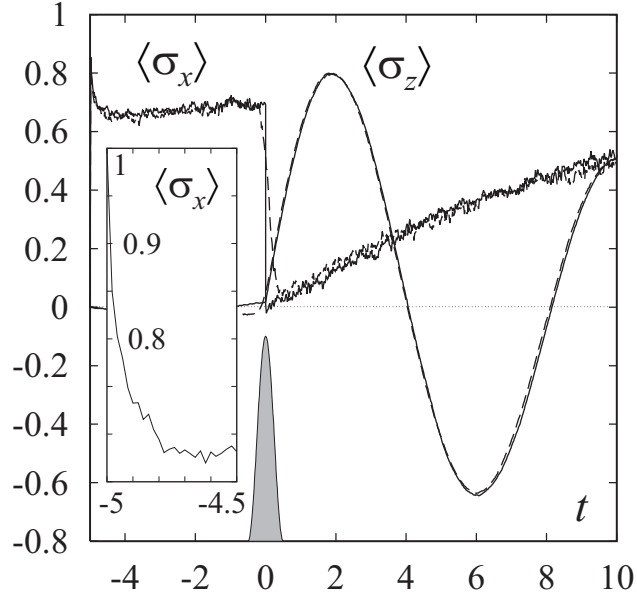}
    \caption[]{Dynamical simulation with a symmetric superposition as an
    initial state (factorizing with the reservoir state). The relaxed
    state after a rapid initial dephasing (inset) is probed by a
    pulsed field in $z$ direction at $t=0$, which rotates the spin
    vector by an angle of $\pi/2$. Subsequently the residual coherence
    of the relaxed state reveals itself through oscillatory
    dynamics. The rise of $\langle \sigma_z\rangle$ {\em above} the
    relaxed value of $\langle\sigma_x\rangle$ is a counterintuitive
    observation, indicating that the quantity $\langle\sigma_x\rangle$
    underestimates the coherence present in the interacting system.
    Solid lines indicate dynamics for a delta pulse at $t=0$, dashed
    lines a finite-width pulse with equal area and shape as indicated in
    the figure.  Parameters are $\Delta = 1$, $\omega_c = 250$, $\alpha
    = 0.05$ and $k_{\rm B}T = 0$.}
    \label{fig:rescoh}
  \end{center}
\end{figure}

In the case of moderate damping $\alpha < 1/2$ and low enough
temperature, the initial rapid decay of a coherent superposition leads
to a relaxed state with some residual coherence, which can be probed by
time-dependent external fields. As is commonly done in elementary
treatments of magnetic resonance, the reduced density matrix of the
two-level system can be parameterized by a polarization or spin vector
$\vec M$,
\begin{equation}
  \label{eq:spinvec}
  \rho = {1\over 2}\left\{
    \left(
      \begin{array}{cc}
        1&0\\ 0&1
      \end{array}
    \right)
    +
    \vec M \cdot \vec \sigma
  \right\} ,
\end{equation}
where $\vec M \equiv \langle \vec \sigma\rangle$. Preparation of the
two-state systems as a symmetric superposition of the eigenstates of
$\sigma_z$ corresponds to a spin vector of unit length pointing in the
$x$ direction. Allowing the system to evolve from a factorized initial
state (as in the previous subsection), the direction of $\vec M$ is
conserved due to symmetry, while the reduction of its length indicates
dephasing. In the case of weak damping, however, the initial dephasing
saturates at a finite value of $\langle\sigma_x\rangle$. As shown in
Fig. \ref{fig:rescoh}, this residual coherence can be probed through a
pulsed external field in the $z$ direction, which changes the relative
phase of the superposition, or, in other words, rotates the spin vector
by an angle of $\pi/2$, leaving it antiparallel to the $y$ axis (without
changing its norm). The following free dynamics displays damped Rabi
oscillations resembling a precession of the spin vector.  However, the
norm $|\vec M| = \sqrt{\langle\sigma_x\rangle^2
  +\langle\sigma_y\rangle^2 +\langle\sigma_z\rangle^2}$ of the spin
vector does not fall monotonously, as one might expect for a dissipative
system. After a quarter oscillation period $\vec M$ points in the $z$
direction, with $M_z$ exceeding the relaxed value of $|\vec M|$ before
the onset of oscillations. This indicates {\em hidden coherence} in
the {\em interacting} system, which does not become visible in the
reduced density matrix. Results for a delta-shaped pulse (solid line)
and a finite-width pulse (dashed line), whose shape is indicated at the
bottom of Fig.  \ref{fig:rescoh}, are virtually identical.

\begin{figure}[tb]
  \begin{center}
    \epsfxsize=0.6\columnwidth
    \epsffile{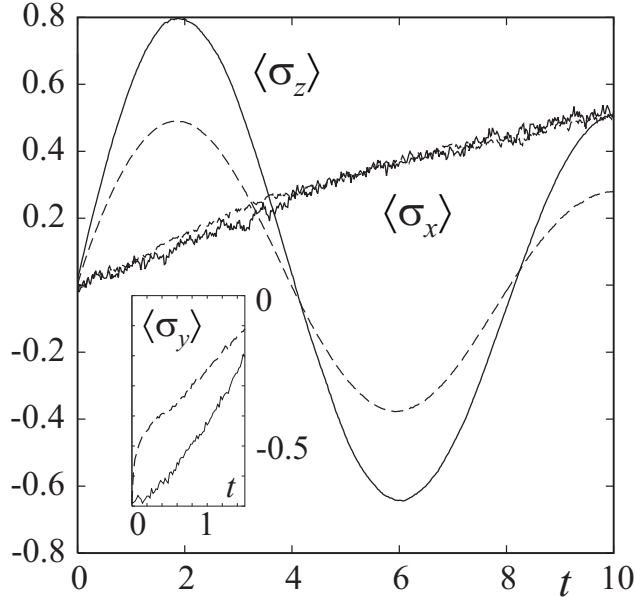}
    \caption[]{Comparison of coherent oscillations starting from a
      relaxed delocalized state (solid lines, same as Fig.
      \ref{fig:rescoh}) and a factorized initial state at $t=0$ with
      identical reduced density matrix $\rho$ (dashed lines). A rapid
      initial decay of $\langle\sigma_y \rangle$ is observed for the
      factorized initial state, but not for the relaxed state, resulting
      in a significant difference in the amplitudes of coherent
      oscillations, while the expectation value $\langle\sigma_x\rangle$
      seems to be unaffected by the choice of initial preparation.
      Parameters are $\Delta = 1$, $\omega_c = 250$, $\alpha = 0.05$ and
      $k_{\rm B}T = 0$.}
    \label{fig:facty}
  \end{center}
\end{figure}

To test this interpretation in terms of hidden coherence, the free decay
of $\langle \sigma_z\rangle$ and $\langle\sigma_y\rangle$ is compared in
Fig. \ref{fig:facty} for the two different preparations of the initial
state at $t=0$. The rotated relaxed state (delta-pulse case of Fig.
\ref{fig:rescoh}) is compared to the dynamics starting with a {\em
  factorized} initial state, where the reduced density matrix at $t=0$
is identical in both cases. A significant preparation effect is found:
In the factorized case, coherence appears to be drastically reduced
compared to the initial preparation derived from a relaxed state,
supporting the notion of hidden coherence. While a rapid initial
dephasing is again observed for the factorizing initial state, it
appears to be absent for the relaxed preparation (see inset of
Fig. \ref{fig:facty}).

\begin{figure}[tb]
  \begin{center}
    \epsfxsize=0.6\columnwidth
    \epsffile{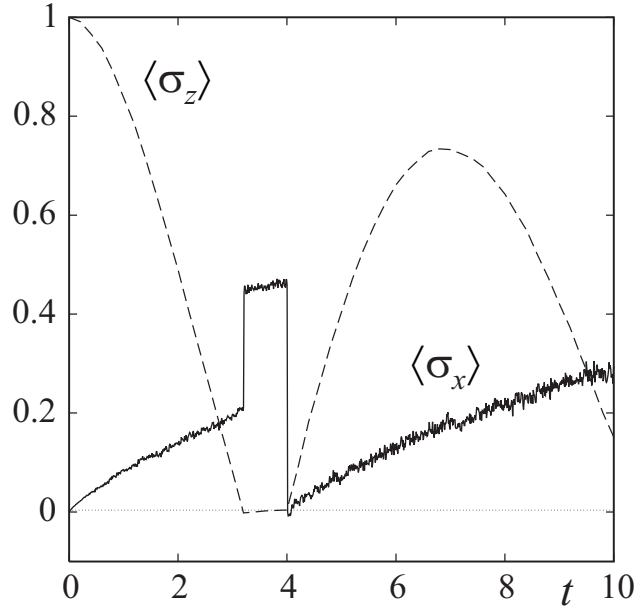}
    \caption[]{Two-pulse dynamics revealing hidden coherence: After the
      free precession of the spin vector starting from a conventional
      localized/factorizing initial state, a pulsed field applied at
      $t=3.2$ rotates the spin vector from the $x-y$ plane onto the $x$
      axis. At $t=4$, a second pulse rotates the spin vector onto the
      $y$ axis, allowing the free precession to resume. The amplitude of
      the following oscillations demonstrates that the value of
      $\langle\sigma_x\rangle$ observed between pulses significantly
      underestimates coherence. Parameters are $\Delta=1$, $\alpha=0.1$,
      $\omega_c/\Delta = 1000$ and $k_{\rm B}T = 0$.}
    \label{fig:twopulse}
  \end{center}
\end{figure}

Signatures of hidden coherence can be observed without taking a
symmetric state for a starting point. Fig. \ref{fig:twopulse} shows a
simulation starting from a conventional factorized preparation involving
a localized state, which has been commonly used as a standard in the
spin-boson literature. After a quarter-period of free oscillatory
dynamics, a pulsed field in $z$ direction rotates the spin vector to
align it with the $x$ direction, effectively stopping the precession of
$\vec M$. A second pulse later aligns the spin vector with the $y$
axis, allowing the system to resume its oscillatory dynamics. Again, the
maximum of $\langle\sigma_z\rangle$ exceeds the previous norm of the
spin vector in the delocalized state between the two pulses.

In this last example, the separation of timescales between system and
reservoir is so pronounced that the linear SLN equation
(\ref{eq:SB:SLN:lin}) with the alternative noise spectrum
(\ref{eq:ndnoise}) can be used for effective simulations. The
simulations discussed in this subsection have also demonstrated that the
SLN approach is capable of performing simulations whose characteristic
timescales differ by several orders of magnitude.

\section{Conclusions and outlook}

A bundle of new numerical simulation strategies is added to the recently
introduced stochastic Liouville-von Neumann equations for open quantum
systems. First, judicious use has been made of the freedom these
equations allow in the choice of complex noise spectra. The resulting
noise spectra largely avoid driving the system resonantly, leading to
markedly improved numerical performance. Additionally, the normalization
of subensembles instead of individual samples improves the stability of
the method in a crucial way. Finally, use is made of the separation of
timescales between dynamical response (inverse bandwidth) and the thermal
timescale, yielding stable and efficient simulations for arbitrarily
long times.

Using the spin-boson model as a test bed, these simulation methods have
been validated as an efficient approach to the dynamics of open quantum
systems. Several examples demonstrate its long-time stability as well as
its suitability for problems with disparate timescales extending over
several orders of magnitude.

A series of simulation results using pulsed fields to extract
information on coherent superpositions suggests that the off-diagonal
elements of the reduced density matrix are not a suitable measure of
quantum coherence in the spin-boson problem; there appears to be hidden
coherence related to the system-reservoir interaction.

Since the stochastic Liouville-von Neumann equations discussed here can
be factorized into Schr\"odinger equations, they can be applied to
larger systems without any complications other than those present
already for conservative systems. The simple expressions found for the
stochastic force operators supports this prospect.

\begin{ack}
  It is a pleasure to acknowledge fruitful exchanges on the present
  topic with Ulrich Weiss, as well as his longstanding commitment as a
  teacher, mentor and friend. Vital discussions with Hermann Grabert
  have also helped shape this work. Support of Deutsche
  Forschungsgemeinschaft under SFB 382 is gratefully acknowledged.
\end{ack}

\section*{Appendix: Dynamical transformation of the probability measure}

In order to transform the linear dynamics of Eq. (\ref{eq:slnlin}) into
the nonlinear norm-conserving form (\ref{eq:slnnorm}) we need to find
a set of functions $\xit(t')$, $\xist(t')$, $\nut(t')$, $\nust(t')$ such
that
\begin{equation}
  \label{eq:Wequiv}
  W[\xit,\xist,\nut,\nust]
  =
  W[\xi,\xis,\nu,\nus] 
  \exp\left(i\int\limits_0^t \d t' \nu(t') \bar r_{t'}\right)
\end{equation}
{\em and} show that the Jacobian determinant of the variable change to the
new functions as independent variables of the noise averaging is unity.

Using vector notation ${\bf z} = (\xi,\xis,\nu,\nus)$, the
probability density $W$ can be formally written as
\begin{equation}
  \label{eq:w:explicit}
  W[\xi,\xis,\nu,\nus] = {1\over N}
  \exp\left(
    -{1\over 2} \int \d t' \int \d t''
    {\bf z}(t')
    {\sf M}(t'-t'')
    {\bf z}^{\rm t}(t'') ,
  \right)
\end{equation}
where $N$ is a normalization constant. Here the superscript ${\rm t}$
denotes a transposed matrix (not a hermitean adjoint) because the inner
product of the underlying space of real functions must be used. The
four-by-four matrix ${\sf M}(t'-t'')$ is related to the noise
correlation matrix $\langle {\bf z}^{\rm t}(t) {\bf z}(t'')\rangle$
through
\begin{equation}
  \label{eq:Mandnoise}
  \int \d\tau {\sf M}(t'-\tau)
  \langle {\bf z}^{\rm t}(\tau)
  {\bf z}(t'')\rangle
  =
  \delta(t'-t'') ,
\end{equation}
i.e., ${\sf M}(t'-t'')$ is the inverse of the noise correlation matrix.
With Eqs. (\ref{eq:w:explicit}) and (\ref{eq:Mandnoise}) it is easy to
show that Eqs. (\ref{eq:xishift}) -- (\ref{eq:nustarshift}) `complete
the square' in the exponent of $W$.

The additional task of computing the corresponding Jacobian only
requires a fairly short calculation. Due to the properties required of
the guide trajectory, one finds the simplification
\begin{equation}
  \label{eq:jacobian}
  J =
  \left|
    \begin{array}{cccc}
      {\delta\xit\over\delta\xi} &
      {\delta\xit\over\delta\xis} &
      {\delta\xit\over\delta\nu} &
      {\delta\xit\over\delta\nus} \\
      {\delta\xist\over\delta\xi} &
      {\delta\xist\over\delta\xis} &
      {\delta\xist\over\delta\nu} &
      {\delta\xist\over\delta\nus} \\
      {\delta\nut\over\delta\xi} &
      {\delta\nut\over\delta\xis} &
      {\delta\nut\over\delta\nu} &
      {\delta\nut\over\delta\nus} \\
      {\delta\nust\over\delta\xi} &
      {\delta\nust\over\delta\xis} &
      {\delta\nust\over\delta\nu} &
      {\delta\nust\over\delta\nus}
    \end{array}
  \right|
  =
  \left|
    \begin{array}{cccc}
      {\delta\xit\over\delta\xi} &
      0&
      {\delta\xit\over\delta\nu} &
      0\\
      {\delta\xist\over\delta\xi} &
      1&
      {\delta\xist\over\delta\nu} &
      0\\
      0&
      0&
      1&
      0\\
      {\delta\nust\over\delta\xi} &
      0&
      {\delta\nust\over\delta\nu} &
      1
    \end{array}
  \right|
  =
  \left|
    {\delta\xit\over\delta\xi}      
  \right| .
\end{equation}
Here the second and fourth column simplify due to conditions
(\ref{eq:anlyxic}) and (\ref{eq:anlynuc}) while the third row reflects
the fact that $\nu$ is not shifted.

Now the definition (\ref{eq:xishift}) of $\xit$ immediately leads to
\begin{equation}
  \label{eq:jacobian2}
  {\delta\xit(t')\over\delta\xi(t'')} = \delta(t'-t'')
  + \int_0^t \d s \chi_R(t'-s) {\delta \bar{r}(s)\over\delta
    \xi(t'')} .
\end{equation}
Because $\chi_R(t'-s)$ is a causal response function and
$\delta\bar{r}(s)\over\delta \xi(t'')$ vanishes for $t''\geq s$ by
construction, the integrand is nonzero only for $t''<t'$. The integral
term in Eq. (\ref{eq:jacobian2}) therefore does not enter the
determinant, and we find
\begin{equation}
  \label{eq:jacobian3}
  J = 
  \left|
    {\delta\xit\over\delta\xi}      
  \right|
  = 1 .
\end{equation}


\begin{thebibliography}{10}

\bibitem{chakr84}
S. Chakravarty and A.~J. Leggett, Phys. Rev. Lett. {\bf 52},  5  (1984).

\bibitem{grabe84}
H. Grabert, U. Weiss, and P. H\"anggi, Phys. Rev. Lett. {\bf 52},  2193
  (1984).

\bibitem{legge87}
A.~J. Leggett {\it et~al.}, Rev. Mod. Phys. {\bf 59},  1  (1987).

\bibitem{weiss99}
U. Weiss, {\em Quantum Dissipative Systems}, 2nd edition ed. (World Scientific,
  Singapore, 1999).

\bibitem{legge86}
A.~J. Leggett,  in {\em Directions in condensed Matter Physics}, edited by G.
  Grinstein and G. Mazenko (World Scientific, Singapore, 1986), Vol.~1, p.\
  187.

\bibitem{esqui98}
{\em Tunneling in Solids}, edited by P. Esquinazi (Springer, Berlin, 1998).

\bibitem{makhl01}
Y. Makhlin, G. Sch\"on, and A. Shnirman, Rev. Mod. Phys. {\bf 73},  357
  (2001).

\bibitem{chior03}
I. Chiorescu, Y. Nakamura, C.~J. P.~M. Harmans, and J.~E. Mooij, Science {\bf
  299},  1869  (2003).

\bibitem{fendl96}
P. Fendley, F. Lesage, and H. Saleur, J. Stat. Phys. {\bf 85},  211  (1996).

\bibitem{baur02}
H. Baur, A. Fubini, and U. Weiss, cond-mat/0211046v1 (unpublished).

\bibitem{schmi83}
A. Schmid, Phys. Rev. Lett. {\bf 51},  1506  (1983).

\bibitem{kane92}
C.~L. Kane and M.~P.~A. Fisher, Phys. Rev. Lett. {\bf 68},  1220  (1992).

\bibitem{stock03}
J.~T. Stockburger, Phys. Stat. Solidi {\bf 237},  146  (2003).

\bibitem{feynm63}
R.~P. Feynman and F.~L. Vernon, Ann. Phys. (N.Y.) {\bf 24},  118  (1963).

\bibitem{egger94}
R. Egger and C.~H. Mak, Phys. Rev. B {\bf 50},  15210  (1994).

\bibitem{leung95}
K. Leung, R. Egger, and C.~H. Mak, Phys. Rev. Lett. {\bf 75},  3344  (1995).

\bibitem{muhlb03}
L. M\"uhlbacher and R. Egger, J. Chem. Phys. {\bf 118},  179  (2003).

\bibitem{makar94}
D.~E. Makarov and N. Makri, Chem. Phys. Lett. {\bf 221},  482  (1994).

\bibitem{sim96}
E. Sim and N. Makri, Chem. Phys. Lett. {\bf 249},  224  (1996).

\bibitem{thorw00}
M. Thorwart, P. Reimann, and P. H\"anggi, Phys. Rev. E {\bf 62},  5808  (2000).

\bibitem{diosi98}
L. Di\'osi, N. Gisin, and W.~T. Strunz, Phys. Rev. A {\bf 58},  1699  (1998).

\bibitem{stock98}
J.~T. Stockburger and C.~H. Mak, Phys. Rev. Lett. {\bf 80},  2657  (1998).

\bibitem{strun99}
W.~T. Strunz, L. Di\'osi, and N. Gisin, Phys. Rev. Lett. {\bf 82},  1801
  (1999).

\bibitem{stock99}
J.~T. Stockburger and C.~H. Mak, J. Chem. Phys. {\bf 110},  4983  (1999).

\bibitem{stock01}
J.~T. Stockburger and H. Grabert, Chem. Phys. {\bf 268},  249  (2001).

\bibitem{stock02}
J.~T. Stockburger and H. Grabert, Phys. Rev. Lett. {\bf 88},  170407  (2002).

\bibitem{perci98}
I. Percival, {\em Quantum state diffusion} (Cambridge University Press,
  Cambridge (U.K.), 1998).

\bibitem{breue02}
H.-P. Breuer and F. Petruccione, {\em The theory of open quantum systems}
  (Oxford University Press, Oxford, 2002), p.\ 625.

\bibitem{kubo63}
R. Kubo, J. Math. Phys. {\bf 4},  174  (1963).

\bibitem{klein95}
H. Kleinert and S.~V. Shabanov, Phys. Lett. A {\bf 200},  224  (1995).

\bibitem{weiss87}
U. Weiss, H. Grabert, and S. Linkwitz, J. Low Temp. Phys. {\bf 68},  213
  (1987).

\bibitem{walls94}
D.~F. Walls and G.~J. Milburn, {\em Quantum Optics} (Springer, Berlin, 1994).

\bibitem{schac91}
R. Schack and A. Schenzle, Phys. Rev. A {\bf 44},  682  (1991).

\bibitem{gardi00}
C.~W. Gardiner and P. Zoller, {\em Quantum noise: a handbook of {M}arkovian and
  non-{M}arkovian quantum stochastic methods with applications to quantum
  optics}, Vol.~56 of {\em Springer series in synergetics}, 2nd ed. (Springer,
  Berlin, 2000).

\bibitem{braun01}
D. Braun, F. Haake, and W.~T. Strunz, Phys. Rev. Lett. {\bf 86},  2913  (2001).

\end{thebibliography}
\end{document}